 \documentclass[page-classic]{epl2} 
\usepackage{epstopdf}
\title{Hadron production in heavy relativistic systems}
\shorttitle{Hadron production} %Insert here a short version of the title if it exceeds 70 characters

\author{Rolf Kuiper\inst{1,2} \and Georg Wolschin\inst{1}}
\shortauthor{R. Kuiper and G. Wolschin}

\institute{  
  \inst{1}                   
 Institut f{\"ur} Theoretische 
Physik, 
        Philosophenweg 16,  
        69120 Heidelberg, Germany\\
          \inst{2} now at Max Planck Institute for Astronomy,
          K\"onigstuhl 17, 69117 Heidelberg, Germany

}
\pacs{25.75.-q}{Relativistic heavy-ion collisions}
\pacs{24.60.Ky}{Fluctuation phenomena}
\pacs{24.10.Jv}{Relativistic models}

\abstract{
We investigate particle production in 
heavy-ion collisions at RHIC energies 
as function of incident energy, and centrality in a 
three-sources Relativistic
Diffusion Model. Pseudorapidity distributions of 
produced charged hadrons in Au + Au and Cu + Cu collisions
at $\sqrt{s_{NN}}$ = 19.6 GeV, 62.4 GeV, 130 GeV and 200 GeV
show an almost equilibrated midrapidity source that tends to increase 
in size towards higher incident energy, and
more central collisions. It may indicate quark-gluon 
plasma formation prior to hadronization. }

\begin{document}

\maketitle

\section{Introduction}
The precise calculation and prediction of transverse momentum
and rapidity distributions of produced particles is of basic
importance in relativistic heavy-ion physics. In this Letter we
propose nonequilibrium-statistical 
methods \cite{wol99} to investigate analytically 
the gradual thermalization in rapidity space occuring in the
course of particle production at the highest available energies. The 
approach is tailored to identify the fraction of produced particles in local 
thermal equilibrium from their pseudorapidity distribution functions 
in heavy systems, with a focus on Cu + Cu and Au + Au.
It may yield indirect evidence for the extent and energy 
dependence of a locally equilibrated parton plasma. 

There exist other theoretical approaches that allow to 
compute rapidity distribution functions for produced particles,
albeit with less precision.
Some of them are based on QCD, such as calculations within the framework of 
the Parton Saturation Model \cite{nar05}.
Ideal hydrodynamics is well developed in its
applications to relativistic collisions, but more realistic
dissipative hydrodynamic approaches are still in the 
early stage of theoretical
development \cite{ko06}.

Thermal models are outstanding in their ability to correctly predict particle 
abundance ratios at midrapidity, or momentum integrated  \cite{pbm01,bec02}. 
But since these approaches do not deal with 
nonequilibrium-statistical effects, one can not expect precise
results for distribution functions whenever 
hadronic or partonic thermalization processes
through multiple collisions on an event-by-event basis
are important \cite{bec02}. Within a thermal model, such
effects could be simulated to some extent by different values
of local temperature and chemical potential when investigating particle 
production at different rapidities. 

Hence, nonequilibrium statistics is the natural choice for a detailed
description of the gradual approach to statistical equilibrium 
in relativistic collisions of heavy systems.
Our Relativistic Diffusion Model (RDM) underlines the nonequilibrium-statistical
features of high-energy heavy-ion collisions, but it also encompasses
kinetic (thermal) equilibrium of the system for times that are
sufficiently larger than the relaxation times of the relevant
variables.

It is of particular interest in relativistic collisions of
heavy systems to determine the fraction of produced particles
that attains - or comes very close to - statistical equilibrium
with respect to a specific macroscopic variable, such as rapidity.
In the three-sources RDM, these are the particles produced in
the midrapidity source. Hence we analyze Au + Au and Cu + Cu 
pseudorapidity distributions of produced particles
at RHIC energies from $\sqrt{s_{NN}}$=19.6 - 200 GeV
corresponding to
% absolute values of the 
beam rapidities
$y_{max}$=3.04 - 5.36.
We determine the transport coefficients and numbers of
produced particles in the midrapidity source as functions
of the incident energies, and centralities. 
%\begin{figure}
%\includegraphics[width=6.6cm]{fig1.eps}
%\caption{Diffusion-model parameters as 
%functions of centrality (mean impact parameter 
%$<b>$)
%for Au + Au at three incident energies $\sqrt{s_{NN}}$ = 19.6, 130 
%and 200 GeV. The time parameter is $\tau_{int}/\tau_{y}$, the width
%of the peripheral sources is $\Gamma_{1,2}$, the width of the
%midrapidity source is $\Gamma_{eq}$, and the corresponding percentage
%of the number of particles produced in this source
%is $n_{eq}$. The size of the midrapidity source increases
%with incident energy, and 
%towards more central collisions.
%Only two sources are required at the lowest energy.}
%\label{fig1}
%\end{figure}
\section{Relativistic Diffusion Model}
In the Relativistic Diffusion Model, the
rapidity distribution of produced particles at RHIC energies
emerges from an incoherent superposition of the beam-like 
components that are broadened in rapidity space through
diffusion processes, and a
near-equilibrium (thermal) component at midrapidity
that may indicate local quark-gluon plasma (QGP) formation.

The time evolution of the distribution
functions is governed by a Fokker-Planck
equation (FPE) in rapidity space
\cite{wol99,wol03,alb00,ryb02,biy02,wols06}
\begin{equation}
\frac{\partial}{\partial t}[ R(y,t)]^{\mu}=-\frac{\partial}
{\partial y}\Bigl[J(y)[R(y,t)]^{\mu}\Bigr]+D_{y}
\frac{\partial^2}{\partial y^2}[R(y,t)]^{\nu} 
\label{fpenl}
\end{equation}\\
with the rapidity $y=0.5\cdot \ln((E+p)/(E-p))$.
The rapidity diffusion coefficient $D_{y}$ that contains the
microscopic physics accounts for the broadening of the
rapidity distributions.
The drift $J(y)$ determines the shift of the mean rapidities
towards the central value, and linear and nonlinear 
forms have been discussed.

Here we use $\mu$ = 1 (due to norm conservation)
and $\nu = 2 - q$ with the nonextensivity parameter \cite{tsa88}
$q = 1$ corresponding to the standard FPE,
and a linear drift function
 \begin{equation}
J(y)=(y_{eq}- y)/\tau_{y}
\label{dri}
\end{equation}
with the rapidity relaxation time $\tau_{y}$, and the equilibrium 
value Ê$y_{eq}$ of the rapidity \cite{wol99,wols06}. This is
the so-called Uhlenbeck-Ornstein \cite{uhl30} process, applied to the
relativistic invariant rapidity for the three components  
$R_{k}(y,t)$ ($k$=1,2,3) of the distribution function
in rapidity space
\cite{wol99,wol03,biy02,wol06} 
\begin{equation}
\frac{\partial}{\partial t}R_{k}(y,t)=
\frac{1}{\tau_{y}}\frac{\partial}
{\partial y}\Bigl[(y-y_{eq})\cdot R_{k}(y,t)\Bigr]
+D_{y}^{k}\frac{\partial^2}{\partial^{2} y}
 R_{k}(y,t).
\label{fpe}
\end{equation}
%\newpage
\begin{figure}
\includegraphics[width=7.6cm]{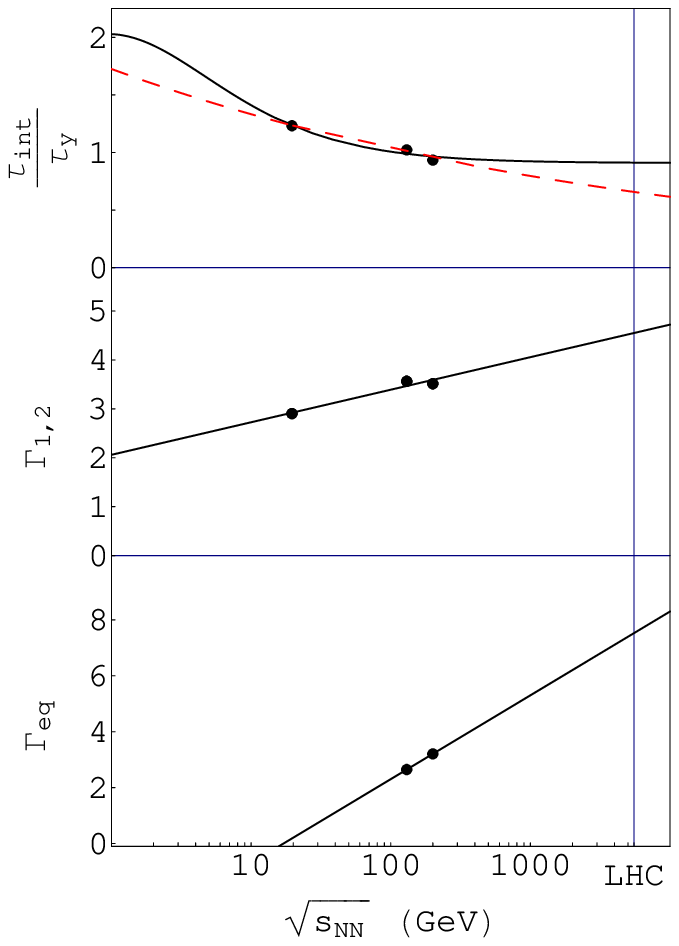}
\caption{Dependence of the Diffusion-Model parameters 
for heavy systems
(central Au + Au at RHIC energies) on the center-of-mass 
energy $\sqrt{s_{NN}}$:  Quotient of interaction time and relaxation 
time for sinh- and exponential (dashed) extrapolation (upper frame);  
width of the peripheral sources including collective expansion (middle
frame); effective width of the midrapidity source (lower frame).
The results are for charged-hadron rapidity distributions,
with extrapolations to LHC energies of 5.52 TeV.
The dots refer to the fit values at $\sqrt{s_{NN}}$=19.6,
130 and 200 GeV.}
\label{fig1}
\end{figure}
\renewcommand{\arraystretch}{1.4}
\begin{center}
\begin{table}
		    \caption{
		    Produced charged hadrons as functions of
centrality in Cu + Cu and Au + Au 
collisions at $\sqrt{s_{NN}}$ = 19.6, 62.4, 130 and 
200 GeV in the Relativistic Diffusion Model.
The number of produced charged particles at each centrality 
is $N_{ch}^{tot}$, the percentage of
charged particles produced in the thermalized source is $n_{ch}^{eq}$.
The ratio $\tau_{int}/\tau_{{y}}$ determines how fast the system 
of produced charged particles equilibrates in rapidity space.
The widths of the peripheral sources are $\Gamma_{1,2}$,
the width of the midrapidity source is $\Gamma_{eq}$.
The $\chi^{2}/d.o.f.$ values with constraints are discussed
in the text.}
\vspace{.2cm}
\label{tab1}
			\begin{tabular}{|c||r|r|r||r|r|r|r|r|r|r||r|}			
			\hline
				\rule[-3mm]{0mm}{8mm}
				%\rule[-3mm]{0mm}{4mm}	
				System &
				$\sigma_{tot}\hspace{4mm}$  &
				$N_{ch}^{tot}$ &
				$n_{ch}^{eq}(\%)$ &
		                  $\frac{\tau_{int}}{\tau_y}$ &$\Gamma_{1,2}$ &$\Gamma_{eq}$ &
				$\frac{\chi^2}{d.o.f.}$\\
				%$\frac{\chi_{min}^2}{d.o.f.}$\\
				\hline\hline
					&0\hspace{1mm}-\hspace{1ex}\hspace{1mm}6\%		& 825		& 3.2	& 1.12	& 3.70& 5.16	& $\frac{4.7}{49}$\\ %	&$\frac{1.9}{49}$	\\
		Cu+Cu		
		&6\hspace{1mm}-\hspace{1mm}15\%				& 681		& 2.2	& 1.11	& 3.70& 4.92	& $\frac{2.4}{49}$\\ %	&$\frac{2.4}{49}$	\\
		62.4 GeV	
		&15\hspace{1mm}-\hspace{1mm}25\%					& 494		& 1.4	& 1.09	& 3.70& 4.73	& $\frac{1.5}{40}$\\ %	&$\frac{0.8}{40}$	\\
					&25\hspace{1mm}-\hspace{1mm}35\%				& 340		& 0.9	& 1.09	& 3.70& 4.57	& $\frac{3.0}{40}$\\ %	&$\frac{0.7}{40}$	\\
					&35\hspace{1mm}-\hspace{1mm}45\%				& 230		& 0.4	& 1.08	& 3.70& 4.45	& $\frac{4.9}{40}$\\ %	&$\frac{0.5}{40}$	\\
		\hline 
					&0\hspace{1mm}-\hspace{1ex}\hspace{1mm}6\%		& 1474		& 7.1	& 1.08	& 4.03& 2.45	& $\frac{2.0}{49}$\\ %	&$\frac{0.9}{49}$	\\
		Cu+Cu		
		&6\hspace{1mm}-\hspace{1mm}15\%				& 1129		& 6.2	& 1.07	& 4.03& 2.40	& $\frac{1.0}{49}$\\ %	&$\frac{0.8}{49}$	\\
		200 GeV		
		&15\hspace{1mm}-\hspace{1mm}25\%				& 791		& 5.4	& 1.06	& 4.03& 2.35	& $\frac{1.8}{49}$	\\ %&$\frac{0.8}{49}$	\\
					&25\hspace{1mm}-\hspace{1mm}35\%				& 536		& 4.9	& 1.05	& 4.03& 2.31	& $\frac{3.5}{49}$\\Ê%	&$\frac{0.8}{49}$	\\
					&35\hspace{1mm}-\hspace{1mm}45\%				& 349		& 4.3	& 1.05	& 4.03& 2.28	& $\frac{5.7}{49}$\\ %	&$\frac{0.8}{49}$	\\
					&45\hspace{1mm}-\hspace{1mm}55\%				& 216		& 4.2	& 1.04	& 4.03& 2.24	& $\frac{8.2}{49}$\\ %	&$\frac{0.7}{49}$	\\
		\hline 
					&0\hspace{1mm}-\hspace{1ex}\hspace{1mm}6\%		& 1691		& -		& 1.23	& 2.90& -		& $\frac{0.7}{28}$\\ %	&$\frac{0.3}{26}$	\\  
		Au+Au		
		&6\hspace{1mm}-\hspace{1mm}15\%				& 1323		& -		& 1.19	& 2.90& -		& $\frac{0.4}{28}$\\ %	&$\frac{0.3}{26}$	\\
		19.6 GeV	
		&15\hspace{1mm}-\hspace{1mm}25\%					& 966		& -		& 1.15	& 2.90& -		& $\frac{0.5}{28}$\\ %	&$\frac{0.2}{26}$	\\
					&25\hspace{1mm}-\hspace{1mm}35\%				& 672		& -		& 1.12	& 2.90& -		& $\frac{1.5}{28}$\\ %	&$\frac{0.3}{26}$	\\	
					&35\hspace{1mm}-\hspace{1mm}45\%				& 429		& -		& 1.10	& 2.90& -		& $\frac{1.9}{27}$\\ %	&$\frac{0.2}{25}$	\\
		\hline 
					&0\hspace{1mm}-\hspace{1ex}\hspace{1mm}6\%		& 4233		& 13.2	& 1.02	& 3.56& 2.64	& $\frac{3.7}{49}$ \\ %	&$\frac{0.8}{49}$	\\
		Au+Au		
		&6\hspace{1mm}-\hspace{1mm}15\%				& 3318		& 11.9	& 1.02	& 3.56& 2.45	& $\frac{1.0}{49}$	\\ %&$\frac{1.0}{49}$	\\
		130 GeV		
		&15\hspace{1mm}-\hspace{1mm}25\%				& 2313		& 10.9	& 1.01	& 3.56& 2.28	& $\frac{1.3}{41}$\\ %	&$\frac{0.4}{41}$	\\
					&25\hspace{1mm}-\hspace{1mm}35\%				& 1559		& 10.0	& 1.01	& 3.56& 2.14	& $\frac{3.1}{41}$	\\ %&$\frac{0.4}{41}$	\\
					&35\hspace{1mm}-\hspace{1mm}45\%				& 1005		& 9.3		& 1.00	& 3.56& 2.05	& $\frac{6.2}{41}$\\ %	&$\frac{0.5}{41}$	\\
					&45\hspace{1mm}-\hspace{1mm}55\%				& 615		& 8.6		& 1.00	& 3.56& 1.93	& $\frac{9.7}{41}$	\\ %&$\frac{0.7}{41}$	\\
		\hline 
					&0\hspace{1mm}-\hspace{1ex}\hspace{1mm}6\%		& 5123	& 26.3	& 0.93	& 3.51& 3.20	& $\frac{1.1}{49}$	\\ %&$\frac{0.7}{49}$	\\  
		Au+Au		
		&6\hspace{1mm}-\hspace{1mm}15\%				& 3987	& 24.8	& 0.93	& 3.51& 3.08	& $\frac{0.8}{49}$	\\ %&$\frac{0.5}{49}$	\\ 
		200 GeV		
&15\hspace{1mm}-\hspace{1mm}25\%				& 
2827	& 23.7	& 0.92	& 3.51& 2.97	& $\frac{2.3}{49}$\\	%&$\frac{0.6}{49}$	\\
					&25\hspace{1mm}-\hspace{1mm}35\%				& 1916	& 22.7	& 0.92	& 3.51& 2.90	& $\frac{7.9}{49}$\\	%&$\frac{0.8}{49}$	\\
					&35\hspace{1mm}-\hspace{1mm}45\%				& 1251	& 21.9	& 0.92	& 3.51& 2.83	& $\frac{11.7}{49}$\\	%&$\frac{1.2}{49}$	\\
					&45\hspace{1mm}-\hspace{1mm}55\%				& 762	& 21.1	& 0.91	& 3.51& 2.76	& $\frac{15.4}{49}$\\	%&$\frac{1.0}{49}$	\\
		\hline

			\end{tabular}		
\end{table}
\end{center}
Since the equation is linear, a superposition of the distribution
functions \cite{wol99,wol03} using the initial conditions
$R_{1,2}(y,t=0)=\delta(y\pm y_{max})$
with the absolute value of the beam rapidities 
$y_{max}$, and $R_{3}(y,t=0)=\delta(y-y_{eq})$  
yields the exact solution. In the solution, the mean values and variances 
are obtained analytically from the moments 
equations. The equilibrium value
$y_{eq}$ is calculated at each centrality from energy- and 
momentum conservation among the participants. For symmetric systems 
it is $y_{eq}=0$ independently of centrality, but its deviation from 
zero is important for precise calculations of rapidity
distributions in case of asymmetric systems
\cite{wols06}.
\section{Pseudorapidity distributions}
If particle identification is not available, one has to
convert the results to pseudorapidity, 
$\eta=-$ln[tan($\theta / 2)]$ with the scattering angle $\theta$.
The conversion from $y-$ to $\eta-$
space of the rapidity density
\begin{equation}
\frac{dN}{d\eta}=\frac{p}{E}\frac{dN}{dy}=
J(\eta,\langle m\rangle/\langle p_{T}\rangle)\frac{dN}{dy} 
\label{deta}
\end{equation}
is performed through the Jacobian
\begin{eqnarray}
\lefteqn{J(\eta,\langle m\rangle/\langle p_{T}\rangle) = \cosh({\eta})\cdot }
\nonumber\\&&
\qquad\qquad[1+(\langle m\rangle/\langle p_{T}\rangle)^{2}
+\sinh^{2}(\eta)]^{-1/2}.
\label{jac}
\end{eqnarray}
We approximate the average mass $<m>$ of produced charged hadrons in the
central region by the pion mass $m_{\pi}$, and use a
mean transverse momentum $<p_{T}>$ = 0.4
GeV/c. Due to the conversion,
the partial distribution functions are different from Gaussians.

\begin{figure}
\includegraphics[width=8.6cm]{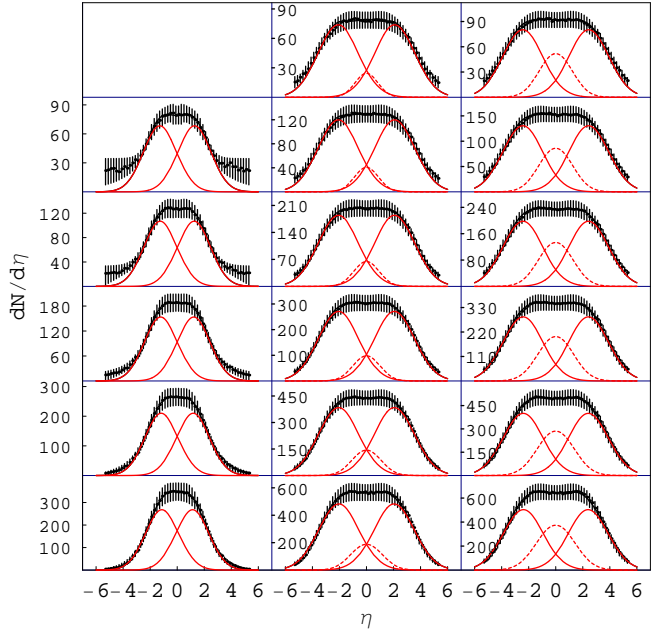}
\caption{Calculated pseudorapidity distributions of 
produced charged particles from
Au + Au collisions at $\sqrt{s_{NN}}$ = 19.6, 130 and 200 GeV for
six collision centralities (sequence as in Figure \ref{fig3}, bottom
frames for central collisions)
in comparison with
PHOBOS data \cite{bac03}. The analytical RDM-solutions are
optimized in a fit to the data. The corresponding 
$\chi^{2}$-values are given in Table \ref{tab1}. Dashed curves show
the midrapidity sources for hadron production.} 
\label{fig2}
\end{figure}

In the linear two-sources version, the Relativistic Diffusion Model had 
been applied to
pseudorapidity distributions of produced charged hadrons in Au+Au 
collisions at RHIC energies of 130 GeV and 200 GeV 
by Biyajima {\it et al.} \cite{biy02}. However, it soon turned 
out from the net-proton results
\cite{wol03},
and from general considerations, that an additional midrapidity 
source is required \cite{wol03,biya04}. This source for particle
production arises mostly from gluon-gluon collisions and emerges at
very short times. In the model we assume that it is generated
at t=0 and $y_{eq}$ with full strength, and then spreads in rapidity space
according to Eq.(\ref{fpe}) during the strong-interaction time.
It comes close to local thermal equilibrium with respect to the
variable rapidity during the interaction time
$\tau_{int}$ and hence, we use the notion $R_{eq}^{loc}(y,t)$ for the
associated partial distribution function in y-space, with 
$N_{ch}^{eq}$ charged particles.  

The validity of this picture was underlined by our 
recent investigation of the d + Au system at 200 GeV 
in the three-sources model \cite{wols06}. Here, an 
accurate modeling of the gradual approach of the system to thermal 
equilibrium in rapidity space was obtained. 
In particular, the dependence of the asymmetric pseudorapidity 
distribution functions on centrality was precisely described.
In the present investigation, however, we concentrate on heavy
symmetric systems where QGP formation is more likely. 

\begin{figure}
\includegraphics[width=7.6cm]{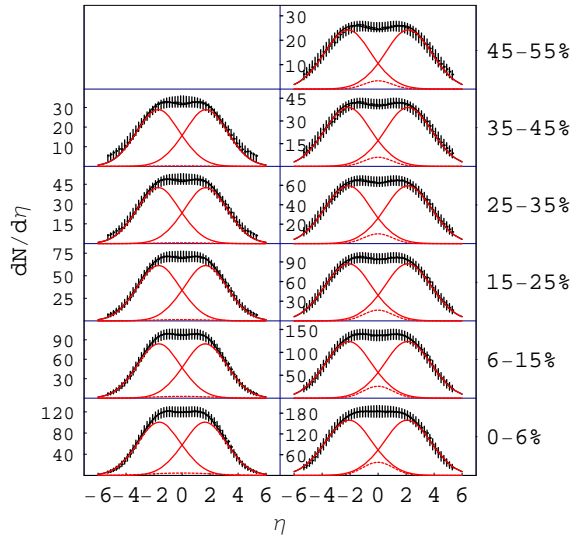}
\caption{Pseudorapidity distributions of charged particles from
Cu + Cu collisions at $\sqrt{s_{NN}}$ = 62.4 and 200 GeV for six 
collision centralities in comparison with
PHOBOS data \cite{nou05}. The analytical diffusion-model solutions are
optimized in a fit to the data, with constraints discussed
in the text. The corresponding 
 $\chi^{2}$-values are given in Table \ref{tab1}.} 
%together with the
%$\chi^{2}_{min}$-values of an unconstrained fit.} 
\label{fig3}
\end{figure}

The dependence of the diffusion-model parameters on incident energy in
central Au + Au collisions at RHIC is displayed in Figure \ref{fig1}.
%, together with extrapolations to LHC energies of 5.52 TeV.
Resulting values for the time parameter
$\tau_{int}/\tau_{y}$ are shown as function of incident energy in
the upper frame, with a functional dependence 
on the absolute value $y_{max}$ of the beam rapidity and hence, on energy given by 
\begin{equation}
\frac{\tau_{int}}{\tau_{y}} 
\propto \frac{y_{max}N_{part}}{\sinh(y_{max})} 
\label{hdely}
\end{equation}
that is discussed in more detail in \cite{kw06},
whereas the dashed curve assumes an exponential dependence
on $\log(\sqrt {s_{NN}})$.
The partial widths as functions of energy within the RHIC 
range for Au + Au are shown in the 
middle and lower frames of Figure \ref{fig1} for both 
peripheral and midrapidity
sources, which differ for produced hadrons.
The widths are effective values: beyond the statistical widths
that can be calculated from a dissipation-fluctuation theorem,
they include the effect of collective expansion.
Here we have plotted the 
values resulting from the $\chi^{2}$-minimization that include the
time evolution up to $\tau_{int}$, including collective expansion 
\begin{equation}
\Gamma_{1,2,eq}=[8 \ln(2) \cdot D_{1,2,eq}^{eff}\cdot \tau_{y}
\cdot (1-\exp(-2\tau_{int}/\tau_{y}))]^{1/2}.
\label{gamh}
\end{equation}

The charged-particle distribution in rapidity space is obtained
as incoherent 
superposition of nonequilibrium and local equilibrium solutions of
 (\ref{fpe}) 
\begin{eqnarray}
    \lefteqn{
\frac{dN_{ch}(y,t=\tau_{int})}{dy}=N_{ch}^{1}R_{1}(y,\tau_{int})}\nonumber\\&&
\qquad\qquad +N_{ch}^{2}R_{2}(y,\tau_{int})
+N_{ch}^{eq}R_{eq}^{loc}(y,\tau_{int})
\label{normloc1}
\end{eqnarray}
with the interaction time $\tau_{int}$ (total integration time of the
differential equation). In the present work, 
%the integration is 
%stopped at the value of 
$\tau_{int}/\tau_{y}$ is determined together with the set
of other free parameters $\Gamma_{1,2,eq}$ and $n_{ch}^{eq}$
from the $\chi^{2}-$minimization
% that produces the minimum $\chi^{2}$
with respect to the data and hence, the
explicit value of $\tau_{int}$ is not needed as an input. 
%The result for central collisions is $\tau_{int}/\tau_{y} 
%\simeq 0.4$, other values are given in Table \ref{tab1}.
The resulting values for $\tau_{int}/\tau_{y}$ are given
in Table \ref{tab1}. Although there are rather rapid changes of $\chi^{2}$
in narrow intervals of $\tau_{int}/\tau_{y}$ due to the simultaneous
dependence on the other parameters, the fitting procedure provides a reliable criterion
for the determination of the diffusion-model parameters.
\section{Heavy systems at RHIC}
Our results for pseudorapidity distributions of produced charged 
hadrons 
at six different centralities and three incident energies in
Au + Au collisions are shown in Figure \ref{fig2}
in comparison with PHOBOS data \cite{bac03}.
Corresponding results for Cu + Cu at two incident energies are
given in Figure \ref{fig3}
compared with preliminary PHOBOS data 
\cite{nou05}. At the lowest energy, only 
two sources are needed for the optimization of the RDM-parameters in 
a $\chi^{2}$-fit, whereas three sources are indeed required at the higher 
energies.
%The percentage of particles in the midrapidity source is
%given in Table \ref{tab1}. 
At the highest energy of 200 GeV,
the Cu + Cu system requires a smaller percentage (7{\%}) of particles in the
midrapidity source compared to Au + Au, where 26{\%} of the produced 
hadrons are in the equilibrated source for central collisions.
This is consistent with the
expectation that heavier systems are more likely to produce a locally
equilibrated quark-gluon plasma.

%The $\chi^{2}$-minimization program has been written in Mathematica 
%for the purpose of this work \cite{kui06}. In a previous 
%investigation \cite{wols06}, the CERN minuit code \cite{jam81} was 
%used. We have verified that for 200 GeV Au + Au, the results are 
%identical. When the execution stops at minimum $\chi^{2}$, the
%values of the time parameter, and of the effective widths of
%the partial distribution functions are determined.

The parameters of these calculations are summarized in Table 
\ref{tab1} 
%. The
%number of particles produced in the three sources, the time parameters 
%and the effective widths of the partial distributions (including the
%time evolution) are shown 
together with $\chi^{2}/d.o.f.$. The number of degrees of 
freedom ($d.o.f.$) is the number of data points minus the number
of free parameters.
%In the $\chi^{2}$-optimization of the diffusion-model parameters
We have not aimed at the absolute minimum of $\chi^{2}$
as in \cite{biy02,wols06}
because this does not provide a sufficiently precise determination of the
free parameters.
Instead, we use physically meanigful constraints to
reduce the number of free parameters, and to determine a local minimum.

In particular, we impose a linear decrease of the time parameter
$\tau_{int}/\tau_{y}$,
of the percentage of particles in the midrapidity source $n_{ch}^{eq}$,
and of
the partial widths $\Gamma_{1,2,eq}$ with increasing impact parameter. 
Using these 
constraints, it is possible to obtain excellent results in the
$\chi^{2}$-optimization.
%as shown in Table \ref{tab1}.
Regarding the
centrality dependence at fixed incident energy, the
increase of the size of the midrapidity source towards more
central collisions provides a good reproduction of the data. 
This is physically reasonable since the midrapidity source
is expected to be more important towards more
central collisions, where it may originate from an
equilibrated parton plasma because of the high energy density.

In a $\chi^{2}-$minimization without any physical constraints 
\cite{wols06}, the results
for the size of the central source as function of centrality
have not shown such a trend. Instead the percentage of hadrons in
the midrapidity source rises for more peripheral collisions, because the
number of charged hadrons produced in the beam-like regions
of pseudorapidity space falls more strongly than the overall
number of produced particles. In view of the good quality
of the constrained centrality dependence in the present work,
however, this particular result which is physically difficult 
to understand may turn out to be an artifact of the
fit procedure.

Based on our present calculations at RHIC energies, we
have also predicted rapidity distributions of 
produced charged hadrons in central Pb + Pb collisions
at LHC energies using the extrapolations displayed in 
Figure \ref{fig1}. The results are shown in \cite{kw06}.

\section{Conclusion}
To conclude, we have described charged-hadron pseudorapidity distributions
in collisions of heavy symmetric systems at four RHIC energies 
and six centralities with high precision in
a three-sources Relativistic Diffusion Model (RDM) for
multiparticle interactions.
Analytical results for the pseudorapidity
distribution of charged hadrons at all investigated
centralities are found to be in 
excellent agreement with the available data. An extrapolation of the
transport parameters to LHC energies has been 
performed.

At the highest RHIC energy of 200 GeV, about 26{\%} of the 
charged hadrons in central Au + Au collisions
are produced in the midrapidity source. These 
particles come very close to statistical equilibrium in rapidity space
during the strong-interaction time.
In central collisions with
very high energy density, they are likely to originate from
a thermalized quark-gluon plasma. The midrapidity source 
is less pronounced towards
more peripheral collisions, lower energies, and in smaller systems
such as Cu + Cu. It vanishes at the lowest energy of 19.6 GeV. \\

\end{document}